\pageno 1

{\ }

\rm

\centerline{\bf The composition of Io - a late confirmation}

\vskip1cm

\centerline{\bf Vladan Celebonovic}

\vskip.5cm

\centerline{\it Institute of Physics,Pregrevica 118,11080 Zemun-Beograd,Yugoslavia}

\centerline{\it e-mail:celebonovic@exp.phy.bg.ac.yu}

\centerline{\it and:vladanc@mrsys1.mr-net.co.yu}

\vskip2cm

\noindent {\bf Abstract:} Nine years ago,working within the framework of 
theoretical dense matter physics,the present author has determined the 
chemical composition of the Galileian satellites of Jupiter.A few months 
ago,in the flyby of the GALILEO space probe,some of the theoretical 
predictions were confirmed.

\vskip1cm

The aim of this note is not to communicate  a new result in astronomy or some related science.It is,instead,to report on a recent experimental confirmation of a calculation made by the  author some years
 ago (Celebonovic,1987).
In the autumn of 1987.,the Astronomical Observatory in Beograd  organized the 
II Workshop "Astrophysics in Yugoslavia". As a contribution to this 
meeting,the present author has theoretically determined the chemical 
composition of the Galileian satellites.

The calculation was performed within the semiclassical theory of dense matter 
proposed by P.Savic and R.Kasanin (Savic and Kasanin, 1962/65). For a recent 
review of this theory see,for example, (Celebonovic,1995a,b) and references 
given there.Predictions of this theory were compared with laboratory high pressure data in (Celebonovic,1992).

Input data needed in astronomical applications of this theory are
the mass and radius of the object under study.As output,one gets the number and radii of the possible layers in the interior,the distribution of  
pressure,density and temperature with depth,the strength of the magnetic 
field,the angular speed of rotation and the chemical composition.The 
composition is obtainde as the value of the mean atomic mass of the mixture of 
materials making up the object.Any such value can be fitted by various 
combinations of chemical substances,and in choosing the real one,
care has to be taken to include any possible existing observational data.

It was reported at the '87 meeting that Io and Europa have similar values of the mean atomic mass,which can be fitted by a mixture of
FeSiO$_3$ + FeS + SO$_2$ + N$_2$H$_4$H$_2$O. No calculation of the internal structure
was undertaken at the time.

The confirmation of these predictions came in June '96 in form of
a  NASA press release dated May 3,1996 (Isbell and Murrill,1996).This
release reported on the flyby of the satellite Io by the GALILEO space probe on December,7,1995.,when the probe passed within 559 miles of the
moon.

The analyses of the perturbations in the probe's orbit during closest
approach were performed by GALILEO'S celestical mechanics team  (Anderson et al.,1996). It was shown that Io's interior consists of two layers.In the center is a metallic core,most probably made up of iron and iron sulfide,and with a diameter of 560 mi
formed either when Io originally formed,or later due to tidal heating of
its interior.

\vfil\eject

     \vskip1cm

{\bf Conclusion}

\vskip.5cm

In this note we have described how a theoretical result obtained
"at the top of a pen" was experimentally confirmed  nearly a decade later
by an ongoing space mission. As a question of principle,this can be interpreted as one more illustration of the correctness of the physical principles
and calculational procedures of the theory of dense matter proposed by
Savic and Kasanin.

     \vskip1cm

References:

          \vskip.5cm

\item{}\kern-\parindent{Anderson,J.et.al.: 1996, Science,{\bf 272}, 709.}

\item{}\kern-\parindent{Celebonovic,V.: 1987,in:II Workshop "Astrophysics in Yugoslavia"
Program and Abstracts published by the Astronomical Observatory
(Ed.by M.S.Dimitrijevic),p.41.}

\item{}\kern-\parindent{Celebonovic,V.: 1992,Earth,Moon and Planets, {\bf 58}, 
203.}

\item{}\kern-\parindent{Celebonovic,V.: 1995a, Bull.Astron.Belgrade, {\bf 
151}, 37.}

\item{}\kern-\parindent{Celebonovic,V.: 1995b,in: Publ.Obs.Astron.Belgrade, 
{\bf 48}, 139.}

\item{}\kern-\parindent{Isbell,D.and Murill,M.B.: 1996,NASA Press Release 
96-89 \ \hfill\ \break (avaliable
at http:// www.spacelink.msfc.nasa.gov)}

\item{}\kern-\parindent{Savic,P.and Kasanin,R.: 1962/65,The Behaviour of Materials Under High
Pressure I-IV, Ed. by SANU, Beograd.}

     \bye